\documentclass[11pt]{amsart}
\usepackage{amssymb}
\usepackage{graphicx}
\usepackage{epstopdf}
 \newtheorem{thm}{Theorem}[section]
 
 \newtheorem{lem}[thm]{Lemma}
 \newtheorem{prop}[thm]{Proposition}
 \theoremstyle{definition}
 
 \theoremstyle{remark}

 \numberwithin{equation}{section}

\newcommand{\CV}{\hbox{{$\mathcal V$}}}
\newcommand{\CI}{\hbox{{$\mathcal I$}}}

\newcommand{\CC}{\hbox{{$\mathcal C$}}}
\newcommand{\CZ}{\hbox{{$\mathcal Z$}}}
\newcommand{\CD}{\hbox{{$\mathcal D$}}}
\newcommand{\CS}{\hbox{{$\mathcal S$}}}
\newcommand{\CE}{\hbox{{$\mathcal E$}}}

\newcommand{\Z}{\mathbb{Z}}
\newcommand{\C}{\mathbb{C}}
\newcommand{\cg}{\hbox{{$\mathfrak g$}}}
\newcommand{\note}[1]{}

\newcommand{\extd}{{\rm d}}

\newcommand{\isom}{{\cong}}
\newcommand{\eps}{{\epsilon}}
\newcommand{\tens}{\mathop{\otimes}}
\newcommand{\id}{{\rm id}}
\newcommand{\<}{\langle}
\renewcommand{\>}{\rangle}
\newcommand{\del}{\partial}
\newcommand{\End}{{\rm End}}
\newcommand{\trace}{{\rm Tr}}

\newcommand{\Dsl}{{\nabla\kern -8pt /}}

\newcommand{\la}{\triangleright}
\renewcommand{\o}{{}_{\scriptscriptstyle(1)}}
\renewcommand{\t}{{}_{\scriptscriptstyle(2)}}

\newcommand{\eqn}[2]{\begin{equation}#2\label{#1}\end{equation}}

\begin{document}

\title[Noncommutative Riemannian geometry]{Algebraic approach to quantum gravity III: noncommmutative Riemannian geometry}
\author{S. Majid}
\address{School of Mathematical Sciences\\
Queen Mary, University of London\\ 327 Mile End Rd,  London E1
4NS, UK\\ \& 
Perimeter Institute for Theoretical Physics\\
31 Caroline St N., Waterloo, ON N2L 2Y5, Canada}
\thanks{The work was completed on leave at the Mathematical Institute, Oxford, UK}

\email{s.majid@qmul.ac.uk}



\maketitle

\begin{abstract}  This is a self-contained introduction to quantum Riemannian geometry based on quantum groups as frame groups, and its proposed role in quantum gravity. Much of the article is about the generalisation of classical Riemannian geometry that arises naturally as the classical limit;  a theory with nonsymmetric metric and a skew version of metric compatibilty. Meanwhile, in quantum gravity a key ingredient of our approach is the proposal that the differential structure of spacetime is something that itself must be summed over or `quantised' as a physical degree of freedom. We illustrate such a scheme for quantum gravity on small  finite sets.  \end{abstract}

\section{Introduction}

Why is quantum gravity so hard? Surely it is because of its nonrenormalisable nature leading to UV divergences that cannot be tamed. However, UV divergences in quantum field theory arise from the assumption that the classical configurations being summed over are defined on a continuum. This is an assumption that is not based on observation but on mathematical constructs that were invented in conjunction with the classical geometry visible in the 19th century.  {\em A priori} it would therefore be more reasonable to expect classical continuum geometry to play a role only  in a macroscopic limit and not as a fundamental ingredient. While this problem might not be too bad for matter fields in a fixed background,  the logical nonsensicality of putting the notion of a classical manifold that is supposed to emerge from quatum gravity as the starting point for quantum gravity inside the functional path integral, is more severe and this is perhaps  what makes gravity special. The idea of course is to have the quantum theory centred on classical solutions but also to take into account nearby classical configurations with some weight. However, taking that as the actual definition is wishful and rather  putting `the cart before the horse'. Note that string theory also assumes a continuum for the strings to move in, so does not address the fundamental problem either.

 If the problems of quantum gravity are indeed such an artifact of a false continuum assumption then what can we do about dropping? Noncommutative geometry (NCG) is a more general, algebraic, framework for geometry that includes the classical continuum case  but goes much beyond it. In the last 20 years it has developed into a systematic computable framework and that is capable of making quantum gravity predictions as well as being a possible basis for a complete theory. 
The `coordinate algebra' here can be noncommutative as it is on a quantum phase space and hence conjecturally on quantum spacetime as well, at least as an effective theory, or it can be finite-dimensional or infinite-dimensional but reflecting a discrete or poor topology of the underlying space. 

Most well-known among frameworks for NCG are Connes' `spectral triples'\cite{Con}. However, the axioms for these are too closely modelled on the classical case or objects very close to classical ones (such as the noncommutative torus or  Heisenberg algebra {\em aka} $\theta$-spacetimes) and do not hold for many other examples of noncommutative geometries such as coming from genuine quantum groups. In this article we cover our alternative  `quantum groups approach' \cite{BrzMa:gau,Ma:rie,Ma:dia,Ma:rieq,Ma:sph} where quantum groups, as the analogues of Lie groups, play a fundamental role not only as key examples but as gauge groups in quantum principal bundles.  Therefore part of this article is a systematic account of quantum bundles.

Note also that whereas  two decades ago most physicists knew only two or three noncommutative algebras (the Heisenberg algebra, symmetry algebras like the angular momentum algebra, and the algebra of all matrices (or all operators on a Hilbert space)), by now it is accepted that noncommutative `algebras' {\em per se} have as rich a structure as that of Riemannian manifolds and indeed as rich a geometric content when geometry is expressed algebraically. To reach this point of acceptance of noncommutative algebras as having their own structure and geometry is perhaps in the long term the most important legacy of quantum groups. These {\em are} algebras and force one to take algebras seriously, while at the same time they are analogues of Lie groups with an analogous geometry. Beyond them are category-theory based `functorial' constructions also coming in part out of quantum groups and their use in constructing 3-manifold invariants but not limited to that. Such methods eventually could be expected to tie up with other approaches to quantum gravity such as spin networks and causal sets that also give up the continuum (but for these there is not yet a fully developed alternative geometry limiting to our familiar continuum one). We will touch upon some aspects of such approaches. 

This article is complementary to \cite{Ma:algII} where we cover the use of NCG in weak-gravity effective theories that might come out of quantum gravity in the form of   noncommutative flat spacetime. This covers  explicit models of `deformed special relativity' and the (several) issues regarding making physical predictions that might be actually tested (such as variable speed of light) and for which there is a large literature on the `quantum gravity phenomenology' side. If there is one important general lesson from these models for the theoretical side, it is that in highly noncommutative cases  there is generally an {\em anomaly} whereby the classical differential calculus cannot be quantised covariantly and forces extra dimensions either in the spacetime or in the Poincar\'e group to neutralise the anomaly.   

By contrast, the present article is concerned with a general formalism for NCG  at the same level as general relativity but with possibly noncommutative coordinate algebras. We will include at least one example with nontrivial cotangent bundle, from \cite{Ma:sph}. If one is of the view that gravity does not need to be quantised as long as it is suitably extended to include quantized matter, then this may be as far as one needs. Alternatively if one assumes that quantum gravity does need a sum over all geometries, NCG allows the geometries inside the summation to be already  more general which is likely needed for self-consistency  and finiteness as explained above. Also in this case, having a better algebraic control of the geometry we can and will do such things as sum over all (noncommutative) differential structures. The article begins with a reprise of these. At the semiclassical level the classical data for a quantum calculus in the symplectic case is a certain type of symplectic connection and we are therefore saying that this is a new field in physics that in conventional terms we should `integrate over'. There is also planned a first article in our  series of three, which will deal with the philosophical basis of the approach as introduced in \cite{Ma:pri}.

\section{Reprise of quantum differential calculus}\label{diff}

The theory we provide throws out conventional topology and analysis, as founded too closely in classical mechanics. Instead, we demand only a unital algebra $A$ over a field $k$. The latter, with care, could be a commutative ring (for example if one were to work over $\Z$ one could safely say that this assumption had been relegated to something unavoidable). For a conventional picture it should be $\C$. 

A differential structure on $A$ means $(\Omega^1,\extd)$ where 
\begin{enumerate}
\item $\Omega^1$ is an $A-A$-bimodule (so one can associatively multiply 1-forms by functions from the left and the right).
\item $\extd: A\to \Omega^1$ obeys the Leibniz rule $\extd(fg)=(\extd f)g+f\extd g$ for all $f,g\in A$.
\item $\Omega^1={\rm span}_k\{f\extd g\}$
\item (Optional connectedness condition) $\ker\extd=k.1$
\end{enumerate}
It is important that this is just about the absolute minimum that one could require in an associative context, but we shall see that it is adequate for Riemannian geometry. 

In general a given algebra can have zillions of differential structures, just as can a topological space in classical geometry; we have to focus on those with some symmetry. To describe symmetry in noncommutative geometry the most reasonable notion is that of a quantum group $H$. Much has been written on such objects and we refer to \cite{Ma:book}. In brief, the minimal notion of a `group multiplication' is, in terms of the coordinate algebra, an algebra homomorphism $\Delta:H\to H\tens H$. We will use the shorthand $\Delta f=f\o\tens f\t$. Associativity of group multiplication corresponds to `coassociativity' of this `coproduct' in the sense $(\Delta\tens\id)\Delta=(\id\tens\Delta)\Delta$. The group identity corresponds to a map $\eps: H\to k$ (it evaluates the function at the group identity in the classical case) characterised by $(\id\tens\eps)\Delta=(\eps\tens\id)\Delta=\id$. Finally, the group inverse appears as an `antipode' map $S:H\to H$ characterised by $\cdot(\id\tens S)\Delta=\cdot(S\tens\id)\Delta=1\eps$. Most relevant for us at the moment, the action of a group on a vector space $V$ appears at the level of $H$ as a `coaction' $\Delta_R:V\to V\tens H$ (here  a right coaction) characterised by
\[ (\Delta_R\tens\id)\Delta_R=(\id\tens\Delta)\Delta_R,\quad (\id\tens\eps)\Delta_R=\id.\]
There is a similar notion for a left coaction. The nicest case for a quantum differential calculus on an algebra is the {\em bicovariant} case when there are both left and right coactions of a Hopf algebra $H$ on $A$ extending to $\Omega^1$ according to 
\[ \Delta_R(f\extd g)=f\o \extd g\o\tens f\t g\t,\quad \Delta_L(f\extd g)=f\o g\o\tens f\t\extd g\t\]
where we use $\Delta_Rf=f\o\tens f\t$, $\Delta_Lf=f\o\tens f\t$ also for the given coaction on $A$ in the two cases.

When a quantum group acts on an algebra we require the coaction to be an algebra homomorphism. In the case of a differential calculus, we require the functions and 1-forms to generate an entire exterior algebra $(\Omega,\extd)$ with  a wedge product, $\extd$ of degree 1 and  $\extd^2=0$, and in the covariant case we require the exterior algebra to be covariant. A lot is known about the construction and classification (by certain ideals) of covariant differential structures on many algebras, but relatively little is known about their extension to an entire exterior algebra; there is a universal extension (basically let the 1-forms generate it and impose the stated requirements) which is always covariant, but  in practice  it gives too large an answer for the geometry and cohomology to be realistic; one has to quotient it further and there will be many ways to do this (so the higher geometry can involve more data).
\medskip

\noindent{\bf Example}\cite{Wor:dif} The quantum group $\C_q[SU_2]$ has a matrix $t^1{}_1=a, t^1{}_2=b$ etc. of four generators with $q$-commutation relations, a unitary $*$-algebra structure and a $q$-determinant relation $ad-q^{-1}bc=1$. This is a quantum group with $\Delta t^i{}_j=t^i{}_k\tens t^k{}_j$.  We take $\Delta_L=\Delta$ (left translation on the quantum group). There is a left-covariant calculus
\[ \Omega^1=\C_q[SU_2].\{e_0,e_\pm\},\quad e_\pm f=q^{\deg(f)}fe_\pm,\quad e_0 f=q^{2\deg(f)}f e_0\]
where $\deg(f)$ the number of $a,c$ minus the number of $b,d$ in a monomial $f$. The $e_\pm,e_0$ will later be a dreibein on the quantum group. At the moment they are a basis of left-invariant 1-forms. Right multiplication on 1-forms is given via the commutation relations shown. The exterior derivative is
\[ \extd a=a e_0+q b e_+,\quad \extd
b=a e_--q^{-2}b e_0,\quad \extd c=c  e_0+q d e_+,\quad \extd d=c
e_--q^{-2}d e_0.\] 
The natural extension to an entire exterior algebra is 
\[ \extd  e_0=q^3 e_+\wedge e_-,\quad
\extd e_\pm=\mp q^{\pm 2}[2]_{q^{-2}}e_\pm\wedge e_0,\quad
(e_\pm)^2=( e_0)^2=0\]
\[
q^2 e_+\wedge e_-+ e_-\wedge e_+=0,\quad  e_0\wedge e_\pm+q^{\pm
4}e_\pm\wedge e_0=0\] where $[n]_q\equiv (1-q^n)/(1-q)$ denotes a
$q$-integer. This means that there are the same dimensions as
classically, including a unique top form $ e_-\wedge e_+\wedge
e_0$. 

{\it \bf Warning:} in most interesting noncommutative examples the requirement of full covariance is so restrictive that a differential structure does not exist with the classical dimensions. For example, for the above quantum group, the smallest bicovariant calculus is 4-dimensional with generators similar to $e_\pm,e_0$ but some different relations, and a new non-classical generator $\Theta$ (see below).

\subsection{Symplectic connections: a new field in physics}

If an algebra for spacetime is noncomutative due to quantum gravity effects, we can suppose it has the form
\[ f\bullet g- g\bullet f=\lambda\{f,g\}+O(\lambda^2)\]
with respect to some deformation parameter $\lambda$, where we write $\bullet$ to stress the noncommutative product. On the right is some Poisson bracket on the manifold $M$ which is supposed to be found in the classical limit. In this sense a Poisson bracket will have to arise in the semiclassical limit of quantum gravity. However, Darboux' theorem says that all nondegenerate (symplectic) Poisson brackets are equivalent so we do not tend to give too much thought to this; we take it in a canonical form. If one does the same thing for a non(super)commutative exterior algebra $\Omega$ it is obvious that one has a graded (super)Poisson bracket. This has been observed by many authors, most recently in the form of a flat connection of some form \cite{Haw,BegMa:sem}.

This is not much use, however, since as we have mentioned there is often no calculus of the correct dimensions for $\Omega$ to be a flat deformation. If one wants to stay in a deformation setting one must therefore, and we shall, temporarily, relax the full axioms of a differential structure. Specifically, we shall suppose them only at the lowest order in $\lambda$ \cite{BegMa:sem}. Now define
\[ f\bullet \tau -\tau\bullet f=\lambda \hat\nabla_f\tau + O(\lambda^2)\]
for all functions $f$ and 1-forms $\tau$. Morally speaking, one can think of $\hat\nabla_f$ as a covariant derivative $\nabla_{\hat f}$ where $\hat f=\{f,\ \}$ is the Hamiltonian vector field generated by a function $f$, and this will be true in the symplectic case. We therefore call it in general a `preconnection'. It further obeys the `Poisson-compatibility' condiition
\[ \hat\nabla_f\extd g-\hat\nabla_g\extd f=\extd\{f,g\}\]
and its curvature (defined in the obvious way) appears in 
\[ [f,[g,\extd h]_\bullet]_\bullet+{\rm cyclic}=\lambda^2 R_{\hat\nabla}(f,g)\extd h+ O(\lambda^3).\]
If the differential structure was strictly associative, this would imply that the jacobiator on the left would vanish and we see that this would correspond to the preconnection being flat. In the symplectic case   the assumption $[f,\omega]=O(\lambda^2)$ (where $\omega$ is the symplectic 2-form) is equivalent to $\nabla$ being a torsion free symplectic connection \cite{BegMa:sem}. So this is the new classical data that should somehow emerge from quantum gravity in the semiclassical limit. If it emerges with zero curvature it might be expected to be the leading part of an associative noncommutative differential calculus, but if it emerges with nonzero curvature, which is surely the  more likely generic case, there will be no noncommutative associative differential structure of classical dimensions. This is exactly what we typically find in NCG.

Without knowing the full theory of quantum gravity, we can also classify covariant  poisson-compatible preconnections in classical geometry, which classifies possibilities in the quantum gravity theory. For example, the theory in \cite{BegMa:sem, BegMa:twi} finds an essentially unique such object for all simple Lie group manifolds $G$ for the type of Poisson-bracket that appears from quantum groups $\C_q[G]$ and it has curvature given by the Cartain 3-form (in the case of $SU_n$, $n>2$ one has a 1-parameter family but also with curvature). Hence if these quantum groups arise from quantum gravity we know the data $\hat\nabla$. In general, however, it is a new field in physics that has to come out of the quantum gravity theory along with the metric and other fields.

\subsection{Differential anomalies and the orgin of time}

When $\hat\nabla$ does not have zero curvature we say that there is a `differential anomaly': even the minimal axioms of a differential structure do not survive on quantisation. We have said that it can typically be neutralised by increasing the dimension of the cotangent bundle beyond the classical one. But then we have cotangent directions not visible classically at all; these are purely `quantum' directions even more remarkable than the unobservable compact ones of string theory.

This is tied up with another natural feature of NCG; in many models there is a 1-form $\Theta$ such that
\eqn{theta}{[\Theta,f]=\lambda\extd f}
where we assume for the sake of discussion a noncommutativity parameter $\lambda$ (it also applies in finite non-deformation cases). This equation has no classical meaning, as both sides are zero when $\lambda\to 0$; it says that the geometry for $\lambda\ne 0$ is sufficiently noncommutative to be `inner' in this sense. Now, if $\Theta$ is part of a basis of 1-forms along with (say) $\extd x_i$  where $x_i$ are some suitable coordinate functions in the noncommutative algebra, we can write $\extd f$ in that basis. The coefficients  in that basis are `quantum partial derivative' operators $\del^i,\del^0$ on $A$ defined by
\eqn{defpar}{ \extd f\equiv (\del^i f)\extd x_i+(\del^0 f)\Theta.}
The quantum derivatives $\del^i$ would have their classical limits if $x_i$ become usual classical coordinates, but $\del^0/\lambda$ will typically limit to some other differential operator which we call the {\em induced Hamltonian}. 

\medskip\noindent{\bf Example}\cite{Ma:tim}: The simplest example is $U(su_2)$ regarded as a quantisation of $su^*$. The smallest connected covariant calculus here is 4D with an extra direction $\Theta$ as well as the usual $\extd x_i$. The associated $\del^0$ operator is
\[ \del^0={\imath \mu \over 2\lambda^2}\left(\sqrt{1+\lambda^2\del_i\del^i}-1\right)\]
where $\mu=1/m$ is a free (but nonzero) parameter inserted (along with $-\imath$) into the normalisation of $\Theta$. If the effect is a quantum gravity one we would expect $\mu\sim\lambda$ as in (\ref{theta}) but in principle the noncommutativity scales of the algebra $A$ and its calculus are independent.  More details of this `flat space' example are in \cite{Ma:tim,Ma:algII}. If we go a little further and explicitly adjoin time by extending our algebra by a central element $t$ with  $\Theta=\extd t$, then  we obtain a natural description of Schroedinger's equation for a particle of mass $m$ on the noncommutative spacetime. The induced calculus on $t$ in this model is actually a finite-difference one.

We call this  general mechanism {\em spontaneous time generation}  \cite{Ma:tim} arising intrinsically from the noncommutive differential geometry of the spatial algebra.  In effect, any sufficiently noncommutative differential algebra `evolves itself' and this is the reason in our view for equations of motion in classical physics in the first place. Note that this  is unrelated to `time' defined as the modular automorphism group of a von-Neuman algebra with respect to a suitable state\cite{ConRov}. 

\medskip
\noindent{\bf Example}: For a `curved space' example, we consider the bicovariant calculus on $\C_q[SU_2]$ mentioned in the warning above.
The phenonomeon is this case is just the same (it is in fact a `cogravity' phenomenon dual to gravity and independent of it). Thus, the 4D calculus  \cite{Wor:dif}, which is the smallest bicovariant one here, has basis $\{e_a,e_b,e_c,e_d\}$ and  relations:
\[ e_a
\begin{pmatrix}a&b\\ c&d\end{pmatrix}=\begin{pmatrix}qa&q^{-1} b\\
qc&q^{-1}d\end{pmatrix}e_a\]
\[  [e_b, \begin{pmatrix}a&b\cr c&d\end{pmatrix}]=q\lambda\begin{pmatrix}0&a\cr 0&c\end{pmatrix}e_d,
\quad [e_c, \begin{pmatrix}a&b\cr c&d\end{pmatrix}]=q\lambda\begin{pmatrix}b&0\cr d&0\end{pmatrix}e_a\]
\[ [e_d,\begin{pmatrix}a\cr c\end{pmatrix}]_{q^{-1}}=\lambda \begin{pmatrix}b\cr d\end{pmatrix}e_b,\quad 
[e_d,\begin{pmatrix}b\cr d\end{pmatrix}]_q=\lambda  \begin{pmatrix}a\cr c\end{pmatrix}e_c+q\lambda ^2 \begin{pmatrix}b\cr d\end{pmatrix}e_a,\]
where $[x,y]_q\equiv
xy-qyx$ and $\lambda=1-q^{-2}$. The exterior differential has the inner form
of a graded anticommutator $\extd =\lambda^{-1}[\Theta,\ \}$ where
$\Theta=e_a+e_d$. By iterating the above relations one may compute \cite{GomMa:non,Ma:ric}:
\begin{eqnarray*}
\extd(c^k b^n d^m)\kern-10pt&=&  \lambda^{-1}(q^{m+n-k}-1)\, c^k b^n d^m
e_d+
 q^{n-k+1}[k]_{q^2}\, c^{k-1} b^n\, d^{m+1}\, e_b \nonumber
\\
    &&\kern-20pt + q^{-k-n}\,
(\, [m+n]_{q^2}\, c^{k+1} b^nd^{m-1}+q[n]_{q^2}\,
c^kb^{n-1}d^{m-1})e_c   \nonumber
\\
&&\kern-20pt +\lambda\, q^{-k-m-n+2}(\, [k+1]_{q^2}\, [m+n]_{q^2}\,  c^k b^n d^m
+
        q[n]_{q^2}\, [k]_{q^2}\, c^{k-1}b^{n-1}d^m) e_a
        \nonumber
\\
    &&\kern-20pt +\lambda^{-1}(q^{-m-n+k}-1)\, c^k b^n  d^m e_a
\end{eqnarray*}
using our previous notations. This is in a coordinate `patch' where $d$ is invertible so that $a=(1+q^{-1}bc)d^{-1}$; there are similar formulae in the other patch where $a$ is inverted. Now to extract the `geometry' of this calculus let us change to a basis $\{e_b,e_c,e_z,\Theta\}$ where $e_z=q^{-2}e_a-e_d$. Then the first three become in the classical limit the usual left-invariant 1-forms on $SU_2$. Writing 
\[ \extd f\equiv (\del^bf)e_b+(\del^cf) e_c+(\del^zf) e_z+(\del^0 f)\Theta\]
we compute from the above on $f=c^kb^nd^m$ that
\[ {(2)_q\over q^2\lambda}\del^0f=q^{-k}(k)_q(n)_q c^{k-1} b^{n-1} d^m+\left({k+n+m\over 2}\right)_q\left({k+n+m\over 2}+1\right)_q f\equiv \varDelta_q f\]
where $ (n)_q\equiv {(q^n-q^{-n)}/(q-q^{-1})}$ is the `symmetrized $q$-integer' (so the first term can be written as $q^{-k}\del^c_q\del^b_qf$ in terms of symmetrized $q$-derivatives that bring down $q$-integers on monomials, see\cite{Ma:book} for notations). The right hand side here is exactly a $q$-deformation of  the classical Laplace Beltrami operator $\varDelta$ on $SU_2$. To see this, let us note that this is given by the action of the Casimir $x_+x_-+{h^2\over 4}-{h\over 2}$ in terms of the usual Lie algebra generators of $su_2$ where $[x_+,x_-]=h$. To compute the action of the vector fields for these Lie algebra generators in the coordinate patch  above we let $\del^b,\del^c,\del^d$ denote partials keeping the other two generators constant but regarding $a=(1+bc)d^{-1}$. Then
\[ \del^b={\del\over\del b}+d^{-1}c{\del\over \del a},\quad 
\del^c={\del\over\del c}+d^{-1}b{\del\over \del a},\quad 
\del^d={\del\over\del d}-d^{-1}a{\del\over \del a}\]
if one on the right regards the $a,b,c,d$ as independent for the partial derivations. Left-invariant vector fields are usually given in the latter redundant form as $\tilde x= t^i{}_j x^j{}_k{\del\over\del t^i{}_k}$ for $x$ in the representation associated to the matrix of coordinates. Converting such formulae over for our coordinate system, we find
\[ \tilde h=c\del^c-b\del^b-d\del^d,\quad \tilde x_+=a\del^b+c\del^d,\quad \tilde x_-=d\del^c\]
\[ \Rightarrow\quad  \varDelta=\del^b\del^c+{1\over 4}(c\del^c+b\del^b+d\del^d)^2+{1\over 2}(c\del^c+b\del^b+d\del^d)\]
which is indeed what we have above when $q\to 1$. The other coordinate patch is similar.  It is worth noting that the 4D (braided) Lie algebra (see below) of $q$-vector fields defined by this calculus \cite{Ma:lie}  is irreducible for generic $q\ne 1$ so one sees also from this point of view that one cannot avoid an `extra dimension'.

Also, by the same steps as in \cite{Ma:tim} we can manifest this extra dimension by adjoining a new central variable $t$ with $\Theta=\extd t$. By applying $\extd$ to $[f,t]=0$ we deduce that 
\[ [e_b,t]=\lambda e_b,\quad [e_c,t]=\lambda e_c,\quad  [e_z,t]=\lambda e_z,\quad [\extd t,t]=\lambda \extd t\]
where the last implies again that the induced calculus on the $t$ variable is a finite difference one:
$\extd g(t)={g(t+\lambda)-g(t)\over\lambda}\extd t\equiv (\del_\lambda g)\extd t$, which is the unique form of noncommutative covariant calculus on the algebra of functions $\C[t]$ with its additive Hopf algebra structure. We also deduce that $(\extd f) g(t)=g(t+\lambda) \extd f$ for any $f=f(a,b,c,d)$ and hence compute $\del^i,\del^0$ (where $i=b,c,z$) in the extended calculus from
\[ \extd(fg)\equiv \del^i(fg)e_i+\del^0(fg)\Theta=(\extd f)g+f\extd g=\del^i(f)e_i g(t)+(\del^0 f)\Theta g(t)+ f\del_\lambda g\Theta.\]
In these terms Schroedinger's equation or the heat equation (depending on normalisation) appears  in a natural geometrical form on the spacetime algebra fields $\psi(a,b,c,d,t)$, regarded as $\psi(t)\in \C_q[SU_2]$, as the condition that $\extd \psi$ is purely spatial, i.e. $\del^0\psi=0$. Explicitly, this means
\[ {\psi(t)-\psi(t-\lambda)\over\lambda}+ {q^2\lambda\over (2)_q}\varDelta_q(\psi(t))=0.\]
Finally, one may change the normalisation of $\Theta$ above to introduce the associated mass parameter and also $\imath$ according to the reality structures in the model. 

For completeness, the rest of the bicovariant exterior algebra here is with $e_b,e_c$ behaving like usual forms or Grassmann variables and
\[ e_z\wedge e_c+q^2
e_c\wedge e_z=0,\quad e_b\wedge e_z+q^2 e_z\wedge e_b=0,\quad
e_z\wedge e_z=(1-q^{-4})e_c\wedge e_b.\]
\[  \extd \Theta=0,\quad \extd e_c=q^2e_c\wedge
e_z,\quad \extd e_b=q^2e_z\wedge e_b,\quad \extd
e_z=(q^{-2}+1)e_b\wedge e_c.\]

\section{Classical weak Riemannian geometry}\label{weak}

In this section we will set aside NCG and prepare Riemannian geometry for quantisation by recasting it in a suitable and slightly weaker form. This theory is due to the author in the first half of \cite{Ma:rie}.  The first thing to note is that in classical Riemannian geometry one defines covariant derivatives $\nabla_X$ for the action of vector fields $X$. Similarly the Riemann curvature $R(X,Y)$ as an operator on vector fields and the torsion $T(X,Y)$. But in NCG we work more naturally with 1-forms which are dual to vector fields. Thus we instead think of a  covariant derivative in the same spirit as for a left coaction $\Delta_L$, namely
\[ \nabla:\Omega^1\to \Omega^1\bar\tens\Omega^1\]
where $\bar\tens$ means tensor product over the algebra of (say smooth) functions on the classical manifold $M$ and $\Omega^1$ means $\Omega^1(M)$. The left hand output of $\nabla$ is `waiting' for a vector field $X$ to be evaluated against it, which would give $\nabla_X$ as usual.  Similarly the standard formulae convert over to\cite{Ma:rie}
\eqn{RT}{ R_\nabla=(\id\wedge\nabla-\extd\tens\id)\nabla: \Omega^1\to \Omega^2\bar\tens\Omega^1,\quad T_\nabla=\nabla\wedge-\extd:\Omega^1\to \Omega^2}
where $\wedge$ means to apply the product $\Omega^1\bar\tens\Omega^1\to \Omega^2$ in the indicated place. Thus curvature is a 2-form valued operator on 1-forms and torsion is a measure of the failure of the projection of $\nabla$ to coincide with the exterior derivative.  In our view the structure of  classical Riemannian geometry is much more cleanly expressed in these terms and in a coordinate free manner.

\subsection{Cotorsion and weak metric compatibility} 

So far, these remarks apply for any covariant derivative. In Riemannian geometry we also need a metric. In noncommutative geometry it is not reasonable to assume a naively symmetric metric (for example in our $q$-deformed example it will be $q$-symmetric as dictated by the stringencies of $\C_q[S0_3]$-invariance). Non-symmetric metrics are also suggested in other contexts such as for T-duality in string theory or Poisson-Lie T-duality \cite{BegMa:poi}. The idea is that semiclassical corrections to classical Riemannian geometry may entail antisymmetric terms given, for example, by the Poisson bivector that also has to come out of quantum gravity (see above) and one has to work in this weaker setting to study certain semiclassical phenomena. Therefore our first bit of generalisation is to define a metric as simply a bundle isomorphism $TM\isom T^*M$ or in our sectional terms $\Omega^{-1}\isom \Omega^1$ (the former denotes the space of vector fields) as a module over the algebra of functions. In plain English it means a nondegenerate tensorial bilinear map $g(X,Y)$ on vector fields, or a nondegenerate element $g\in \Omega^1\bar\tens\Omega^1$. The metric is symmetric if $\wedge (g)=0$ and we are free to impose symmetry in this form in NCG if we want. 

Next, we define an adjoint connection $\nabla^*$ such that
\[ X(g(Y,Z))=g(\nabla_X^*Y,Z)+g(Y,\nabla_X Z)\]
or in dual form
\[( \nabla^*\bar\tens\id +\id\bar\tens\nabla)g=0\quad\Leftrightarrow ((\nabla^*-\nabla)\bar\tens\id)g+\nabla g=0\]
where in the 2nd term of the first expression, the left-hand output of $\nabla$ is understood to be positioned to the far left. Note that $\nabla g$ is defined similarly by extending $\nabla$ as a derivation but keeping its left-hand output to the far left (where it could be safely evaluated against a vector field $X$).

\begin{prop} \label{conabla} $\nabla^*$ is a connection and 
\[(R_{\nabla^*}\bar\tens\id)g+(\id\bar\tens R_\nabla)g=0,\quad ( (T_{\nabla^*}-T_\nabla)\bar\tens\id)g=(\nabla\wedge\bar\tens\id-\id\wedge\nabla)g\]
where the left hand (2-form valued) output of $R_\nabla$ is understood to the far left. 
\end{prop}

The direct proof is tedious but elementary and will therefore be omitted. One also has to check that the constructions are indeed well defined over $\bar\tens$ which is not completely obvious. (The result is implicitly proven a different way using frame bundles, see below, in \cite{Ma:rie}.) The quantity $T_{\nabla^*}$ is called the {\em cotorsion} of $\nabla$ and we see that a {\em torsion free and cotorsion free} connection $\nabla$ is the same thing as torsion free and 
\eqn{wcompat}{ (\nabla\wedge\bar\tens\id-\id\wedge\nabla)g=0.}
We call such a connection a `weak Levi-Civita' connection (it is no longer unique). Note that (\ref{wcompat})  can also be written as $(\wedge\bar\tens\id)\nabla g=0$ or in components, $\nabla_\mu g_{\nu\rho}-\nabla_\nu g_{\mu\rho}=0$ and we call this  `weak metric compatibility'. Why do we need this weakening? Quite simply in most NCG examples that have been computed the weaker version of the Levi-Civita condition actually determines $\nabla$ uniquely within some reasonable context, and the $\nabla$ that arise this way simply do not obey the full $\nabla g=0$. So Riemannian geometry in its usual form does not generalise to most key examples but this weaker version does and indeed suffices. We will give some examples later. As is not untypical in NCG (as we saw for the 1-form $\Theta$),  the classical theory is degenerate and in that limit (\ref{wcompat}) does not suffice, and this in our view is why we have grown artificially used to the stronger form. 

\subsection{Framings and coframings}

Next, it is quite well-known that the notion of a vector bundle may be expressed in algebraic terms simply as the sections $\CE$ being a finitely-generated projective module (the Serre-Swann theorem). This is the line taken for example in \cite{Con}. However, it is important in Riemannian geometry that the vector bundles that arise are not simply vector bundles but are associated to a principal frame bundle. This ensures that the spinor bundle, the cotangent bundle etc. all have compatible structures induced from a single connection on the frame frame bundle. It is also important in physics to have the `moving frame' picture for calculations. Our first problem in noncommutative geometry is, if we replace $GL_n$ or $O_n$ by a quantum group, what flavour of quantum group should we use? There are in fact many different types of Hopf algebras that deform even $GL_n$ whereas we should like to have a general and not specific theory. The answer is to generalise the classical notion of frame bundle to allow for a general choice of frame group.

Thus, let $G\subset X\to M$ be a principal $G$-bundle over $M$ (where $G$ acts on $X$ from the right, say) and $V$ a representation of $G$. There is an associated bundle  $E= P\times_G V\to M$ and its sections may be identified as the space $\CE=C_G(X,V)$ of $V$-valued $G$-equivariant functions on $X$. The following lemma is known to experts:
\begin{lem} Bundle maps $E\to T^*M$ are in 1-1 correspondence with $C(M)$-module maps $C_G(X,V)\to \Omega^1$ and these in turn are in correspondence with $\theta\in \Omega^1_{\rm tensorial}(X,V^*)$, i.e. with $G$-equivariant horizontal 1-forms on $X$.
\end{lem}
A proof is in \cite{Ma:rie} in some other notations. One way is easy: given $\theta$ we multiply it by any $V$-valued function on $X$ and evaluate the $V^*$ against the $V$ to obtain a 1-form on $X$ which actually is the pull-back of a 1-form on $M$ due to the equivariance and horizontality assumptions.  Using this lemma we define:

\begin{enumerate}\item A framing of a manifold $M$ is $(G,X,V,\theta)$ such that  $E\isom T^*M$ or equivalently  $\CE\isom \Omega^1$ by the above maps.
\item A framed weak Riemannian manifold structure on $M$ means framed as above and {\em also} framed by $(G,X,V^*,\theta^*)$ (we call this the coframing). 
\end{enumerate}

Here the framing implies that the dual bundle is $E^*\isom TM$. Given this, a coframing (which is similarly equivalent to $E^*\isom T^*M$) is equivalent to $E^*\isom E$ or $TM\isom T^*M$ given the framing. In other words, it is equivalent to a metric $g$ in the generalised sense above. The latter is given explicitly by
\eqn{gtheta}{ g=\<\theta^*,\theta\>\in \Omega^1\bar\tens\Omega^1}
where the angular brackets denote evaluation of $V^*,V$ and where the result is the pull back of forms on $M$ due to equivariance and horizontality of $\theta,\theta^*$. 

Next, any connection $\omega$ on the principal bundle $X$ induces a covariant derivative $D: \CE\to \Omega^1\bar\tens \CE$ on any associated bundle with sections $\CE$. Given the framing isomorphism this becomes $\nabla:\Omega^1\to\Omega^1\bar\tens\Omega^1$. One also finds that $D\wedge\theta$ corresponds to the torsion $T_\nabla$ and that $F(\omega)=\extd\omega+ \omega\wedge \omega$ acting on the sections corresponds to $R_\nabla$. As shown in \cite{Ma:rie}, none of this actually needs the bundle $X$ to be the usual frame bundle, we have instead `abstracted' the necessary properties in the notion of a framing. Similarly, we can regard $\theta^*$ in the role of framing and have an induced $\nabla^*$ with torsion $T_{\nabla^*}$ corresponding to $D\wedge\theta^*$ and an induced curvature $R_{\nabla^*}$ which is merely adjoint to that for $\nabla$. One may easily verify that $\nabla^*$ is adjoint to $\nabla$ in the sense of Proposition~\ref{conabla} with respect to $g$ defined by (\ref{gtheta}).

In the case of a parallelisable manifold (or a local coordinate patch, for example) one does not need to work globally `upstairs' in terms of $\theta,\theta^*$. In this case $\theta$ is equivalent to a basis of 1-forms $\{ e_a\}$ such that  $\Omega^1=C(M).\{e_a\}$ (a basis over $C(M)$) that transform among themselves under the action of $G$, in other words a $G$-covariant $n$-vielbein. The first part means that the sections of the cotangent bundle are a free module rather than the general case of a projective one. Similarly a coframing is the choice of another collection $\Omega^1=\{e^a\}.C(M)$ which we call an `$n$-covielbein' and which transforms in the dual representation. The corresponding metric (given a framing) is $g=e_a\bar\tens e^a$ or in indices, 
\[ g_{\mu\nu}=e_{a\mu} e^a_{\nu}.\]
Our decision not to assume that the metric is symmetric is reflected in the fact that the vielbein and covielbein are treated independently. We can of course write $e^a=\eta^{ab}e_b$ and call $\eta^{ab}$ the frame metric but it need not be of a fixed or symmetric form (it can vary over the manifold), but at each point should be $G$-invariant.

Finally, in order to do the minimum of gravitational physics we need to be able to define the Ricci tensor and scalar. To do this we need to explicate a `lifting map' 
\[ \imath:\Omega^2\to \Omega^1\bar\tens\Omega^1\]
that splits the surjection $\wedge$ going the other way (so that $i\circ\wedge$ is a projection operator on $\Omega^1\bar\tens\Omega^1$). In classical differential geometry this a trivial map since 2-forms are already defined by antisymmetric tensors so that  $i(\alpha^{\mu\nu}\extd x_\mu\wedge \extd x_\nu)=\alpha^{\mu\nu}\extd x_\mu\bar\tens x_\nu$ does the job. Evaluating $i$ against a vector field $X$, it is equivalent to the interior product of vector fields against 2-forms, but as before we prefer to take the `coaction' point of view.  In NCG we will need to specify this data in the course of constructing the 2-forms rather than to take it for granted.  Then $(i\bar\tens \id)R_\nabla:\Omega^1\to \Omega^1\bar\tens\Omega^1\bar\tens\Omega^1$ and we can now take the trace (at least in the local or parallelizable case, but is seems to work in practice globally) by feeding the middle $\Omega^1$ (say) back into the input. This defines 
\[ {\rm Ricci}=\trace (i\bar\tens\id) R_\nabla = i(F_{j})^{a b} e_b\bar\tens f^j\la e_a\in \Omega^1\bar\tens\Omega^1\]
from the general and from the framed points of view. Here $i(F)$ is expressed in the vielbein basis and $\la$ denotes  the action of the Lie algebra $\cg$ of $G$ with basis $\{f^i\}$ on $V={\rm span}_k\{e_a\}$. Notice that ${\rm Ricci}$ does not depend exactly on the metric or covielbein when set up in this way, rather just on the framing which is half way to a metric. Also note that in these conventions the Ricci tensor is minus the usual one. We then define the Ricci scalar by applying $g^{-1}$ as a map and taking the trace $R=\trace g^{-1}{\rm Ricci}$ again at least in the parallelizable case. It is fair to say that a completely abstract and more conceptual picture of the Ricci curvature and scalar is missing both in classical geometry in this setup and in the NCG case, but the above does give more or less reasonable results in examples including nonparallelizable ones. 

The same remarks apply to the Dirac operator where there is a practical definition as follows: we require some other representation $W$ of $G$ such that the associated bundle with sections $\CS$ can serve as the spinor bundle. What is required for this is a bundle map $\Gamma:\Omega^1\to \End(\CS)$ which at least in the parallelizable case is given by a $G$-equivariant map $\gamma:V\to \End(W)$. The latter is our notion of `generalised Clifford structure'. The Dirac operator is defined by  $\Dsl=\circ(\Gamma\bar\tens\id)D$ where $D:\CS\to \Omega^1\bar\tens\CS$ is the covariant derivative from the connection on $X$ and $\circ$ is the application of $\End(\CS)$ on $\CS$. This constructive approach gives reasonable answers in NCG examples including non-parallelizable ones but a  conceptual picture and axiomatization based on it are missing at the time of writing  (it does not usually obey Connes' axioms for a spectral triple, for example, but may obey some generalised version of them).

Note also that the above reformulation and weakening of classical Riemannian geometry is completely symmetric between 1-forms and vector fields or vielbeins and covielbeins up until the Ricci curvature. In other words the basic geometrical set-up is self-dual and should therefore be better adapted to microlocal analysis or local Fourier transform ideas. This is part of our `vision' of Einstein's equation as a self-duality equation, see Section~6, when both sides are understood properly (in NCG for example). At a more practical level it may be interesting to take these weaker axioms  seriously and ask if there are interesting weak classical solutions, such as new kinds of weak black holes etc. At the moment most attention has been given to quantising the geometry which means by intention landing back on a conventional configuration in the classical limit $\lambda\to 0$, rather than this question for classical weak Riemannian geometry.

\section{Quantum bundles and Riemannian structures}

The above reformulation and weakenig of classical Riemannian geometry is now NCG-ready. All notions in the previous section make sense over any algebra $A$ with differential structure.  Moreover, the extra rigidity provided by a high degree of noncommutativity of the geometry tends to compensate for the weakening and yield canonical answers, for example for Riemannian structures on quantum groups and homogeneous spaces. 

First, a quantum bundle over an algebra $A$ means \cite{BrzMa:gau,BrzMa:rev}:
\begin{enumerate}
\item $H$ a Hopf algebra coacting $\Delta_R:P\to P\tens H$ covariantly on an algebra $P$ with $A=P^H\equiv \{p\in P\, |\, \Delta_Rp=p\tens 1\}\subseteq P$ as the fixed subalgebra.
\item Compatible differential structures where $\Omega^1(H)=H.\Lambda^1_H$ is bicovariant,  $\Omega^1(P)$ is $H$-covariant and $\Omega^1(A)=A(\extd A)A\subseteq \Omega^1(P)$.
\item $0\to P\Omega^1(A)P\to \Omega^1(P)\to P\tens\Lambda^1_H\to 0$ 
is exact. 
\end{enumerate}
Here any quantum group left (or bi)-covariant calculus is a free module of the form shown with basic left-invariant 1-forms spanning a vector space $\Lambda^1_H$. This can be explicitly constructed as a quotient of $\ker\eps\subset H$ by some Ad-invariant right ideal $\CI_H$ and should be thought of as (and typically is) the dual of some finite-dimensional (braided) Lie algebra underlying $H$\cite{GomMa:bra}. The compatibility of the calculi includes the stated requirement that the calculus on $P$ restricted to $A$ gives the desired calculus on $A$ and a further condition on the ideal $\CI_H$ under $\Delta_R$ reflecting `smoothness' of this coaction. The third item is key and says that the kernel of the `left-invariant vector fields map' $\Omega^1(P)\to P\tens \Lambda^1_H$ is precisely the space of horizontal forms $P\Omega^1(A)P$, where the former is an infinitesimal version of the coaction $\Delta_R$ and determined by that. If one evaluates $\Lambda^1_H$ against an element $x$ of the (braided) Lie algebra of $H$, one has a map $\Omega^1(P)\to P$ which is the noncommutative vector field generated by the action of $x$. This notion is now about 14 years old and quite well explored by many authors. A special case is to take so-called universal calculi on $A,H,P$ (so for example the ideal $\CI_H=0$) and 
in this case a quantum bundle is equivalent to an algebraic notion of `Hopf-Galois extension'. This is in some sense the `topological' level of  the theory maximally far from classical differential geometry.

Next, if $V$ is a right $H$-comodule we have $\CE=(P\tens V)^H$ the associated bundle in NCG given by the fixed subspace under the tensor product coaction (it can also be set up as a cotensor product). We work directly with the sections as there is not necessarily any underlying classical total space. Then a framed manifold structure on $A$ means $(H,P,V,\theta)$ where we have the above and an equivariant map
\[ \theta:V\to P\Omega^1(A)\]
(the right hand side here is the space of `left-horizontal' forms) such that the induced map $\cdot (\id\tens \theta):\CE\to \Omega^1(A)$ is an isomorphism . Being a map from $V$ is the same as having values in $V^*$ as we had before (at least in the finite-dimensional case of main interest). 

Finally, a coframing means similarly $(H,P,V^*,\theta^*)$ where $\theta^*:V^*\to \Omega^1(A)P$. A framed and coframed algebra is the same as a framed algebra with metric $g$ given as in (\ref{gtheta}). A (strong) connection $\omega$ on $H$, which we call `spin connection' with respect to the framing induces a covariant derivative $D$ on associated bundles and we define $\nabla$ in the same way as before. In the framed and coframed case we also have $\nabla^*$ and we say that the connection is `weakly metric compatible' if the torsion and cotorsion vanish. The latter is equivalent as before to metric-compatibilty in a skew form (\ref{wcompat}). The torsion and curvature are given by the same formulae as before in (\ref{RT}) and correspond to a certain $\bar D\wedge\theta$ and $F(\omega)$ respectively but now on the quantum bundle (this requires certain flatness conditions on $\Omega^2$ but these hold in practice). Similarly for the Dirac operator in terms of equivariant $\gamma:V\to \End(W)$ at least in the parallelizable case or in terms of $\Gamma$ in the general case.

 Note that in the parallelizable case $\omega$ `upstairs' is determined by $\alpha:\Lambda^1_H\to \Omega^1(A)$ which one should think of as a (braided) Lie algebra-valued connection $\alpha=\alpha_if^i$ with $\{f^i\}$ a basis of $\Lambda^{1*}_H\subset H^*$. The torsion and cotorsion equations then become
\[ \extd e_a+\sum \alpha_i\wedge f^i\la e_a=0,\quad \extd e_a+S^{-1}f^i\la e_a\wedge\alpha_i=0 \] while the curvature `downstairs' is
\[ F(\alpha)=\extd \alpha_i + c_i{}^{jk} \alpha_j\wedge\alpha_k\]
where $f^j f^k=f^ic_i{}^{jk}$ expresses the product of $H^*$ (or the coproduct of $H$). 

All of this would be pie in the sky if not for the fact that over the years a lot has been computed and while there are still some rough edges to the abstract theory, it does apply to a wide range of examples. Still more examples are covered by a generalisation to coalgebras in place of Hopf algebras\cite{BrzMa:geo}.

\medskip
\noindent {\bf Example}\cite{Ma:rie} Just to set the scene with something familiar, let $G$ be a compact Lie group with Lie algebra $\cg$. We use $G$ to frame itself, with $X=G\times G$ trivial. This choice of frame group restricts what kind of $\nabla$ can be induced but is adequate for many purposes. We take $\alpha={1\over 2} e$ where $e\in \Omega^1(G,\cg)$ is the Maurer-Cartan form obeying $\extd e+e\wedge e=0$. We let $V=\cg^*$ with the coadjoint action and basis $\{f_a\}$ say, and we take vielbein $e_a=f_a(e)$ given by the components of the Maurer-Cartan form. In short, $e$ defines the framing and $e/2$ defines the spin connection. The torsion vanishes since $De=\extd e+[\alpha,e]=\extd e+{1\over 2}[e,e]=\extd e+e\wedge e=0$ but its curvature $F(\alpha)=-{1\over 4}e\wedge e$ does not. If we take the coveilbein to be related to the vielbein by the Killing form (so the local metric $\eta$ is the Killing form) then the cotorsion automatically also vanishes. The corresponding $\nabla$ is in fact the usual Levi-Civita connection for the Killing metric (i.e. $\nabla g=0$ as it happens) but we have constructed it in a novel way. 

\medskip
\noindent {\bf Example}\cite{Ma:ric} The above example works just as well for quantum groups. For $A=\C_q[SU_2]$ with its 4D bicovariant calculus as studied in Section~2, we take $H=A$ and $P=A\tens A$. There is a Maurer-Cartan form $e$ and a (braided)-Killing form on the 4D cotangent space which is invariant and used to define the coframing as well. Remarkably, this metric is a $q$-deformation of the Minkowksi metric when all the reality constraints are taken into account, i.e. the extra direction $\Theta$ in $\Lambda^1_H$ enters with a negative signature in the limit. We then solve for a spin connection $\alpha$ (it is no longer just $e/2$ but is uniquely determined) and find the resulting torsion free cotorsion free $\nabla$. One finds $\nabla g=O(\lambda)$ (where $q=\exp({\lambda/2})$) but not zero. Finally, the exterior algebra is given by braided-skew symmetrization and hence, as classically, there is a canonical lifting $i$ given from this braiding. The resulting Ricci tensor is
\[ {\rm Ricci}=-{2 q^2\over [4]_{q^2}}(g+{q^4\over 1+q^2}\Theta\bar\tens\Theta).\]
In other words, apart from the mysterious non-classical extra direction $\Theta$, the noncommutative space is `Einstein'. 

\medskip
\noindent {\bf Example}\cite{Ma:rieq} The nice thing is that algebras of functions on finite groups $G$ are also perfectly good Hopf algebras. Even though the algebra $A=\C(G)$ is commutative, its differential calculus is necessarily noncommutative (there is no non-zero classical differential structure on a finite set). The bicovariant calculi have invariant forms $e_a$ labelled  by Ad-stable subsets (e.g. conjugacy classes) in $G$ not containing the group identity element 1. For $S_3$ there is a natural calculus $\Omega^1=\C(S_3)\{e_u,e_v,e_w\}$ labeled by the 2-cycles $u=(12), v=(23), w=(13)$. There is an element $\Theta=\sum_a e_a$ which makes the calculus inner as we said was typical of a strictly noncommutative geometry. The (braided) Killing form comes out as the standard $g=\sum_a e_a\tens e_a$ and again one can solve for torsion free cotorsion free $\alpha$. Under a regularity condition this is unique and given by
\[ \nabla e_u=-e_u\bar\tens e_u - e_v\bar\tens e_w -e_w\bar\tens e_v+ {1\over 3}\Theta\bar\tens\Theta\]
One can verify that $\nabla g\ne 0$  but that (\ref{wcompat}) does hold as it must by construction. Again there is a nontrivial braiding $\Psi$ that defines the exterior algebra and hence a canonical lifting map $i$. The Ricci curvature is then
\[ {\rm Ricci}={2\over 3}(-g+\Theta\bar\tens \Theta).\]
In other words, apart form this mysterious extra dimension we again have an Einstein space with (what would be positive in usual conventions) constant curvature. Indeed, the canonical NCG of $S_3$ is more like quantum $S^3$ (the quantum group $SU_2$). If one does the same thing for the alternating group $A_4$ one has again a unique spin connection and this time ${\rm Ricci}=0$.

\medskip
\noindent {\bf Example}\cite{Ma:sph} To convince the reader we have to give a nontrivial non-parallelizable example. The one that has been fully worked out is $\C_q[S^2]$ the standard quantum sphere. We define on $P=\C_q[SU_2]$ the coaction $\Delta_Rf = f\tens t^{\deg(f)}$ of the Hopf algebra $H=\C[t,t^{-1}]$ with coproduct $\Delta t=t\tens t$. The latter is the coordinate algebra of $S^1$ and we are giving the $q$-analogue of $S^2=SU_2/S^1$. Then $A=\C_q[S^2]\equiv\C_q[SU_2]_0$, the degree zero subspace. It inherits an algebra structure generated by $b_+=cd$, $b_-=ab$, $b_0=bc$ and differential calculus from that of $\C_q[SU_2]$, where we use the 3D calculus given in Section~\ref{diff}. Finally, we need to take a calculus $\Omega^1(H)=\C[t,t^{-1}]\extd t$ with relations $\extd t. t=q^2 t.\extd t$ and $\extd t^n=[n]_{q^2}t^{n-1}$, which is the unique form of noncommutive differential structure on a classical circle for some parameter (here $q^2$). It is known that all this gives a quantum bundle, essentially it was found in \cite{BrzMa:gau} in some form along with a canonical connection $\alpha$, the $q$-monopole. 

Next, a main theorem of \cite{Ma:sph} is that any quantum homogeneous bundle built from quantum groups $\pi:P\to H$ like this makes the base algebra $A$ a framed quantum manifold. Here
\[ V={P^+\cap A\over \CI_P\cap A},\quad \Theta(v)=S\tilde v\o\extd \tilde v\t,\quad \Delta_R(v)=\tilde v\t\tens S\pi(\tilde v\o)\]
where $\CI_P$ is the ideal that describes the left-covariant $\Omega^1(P)$ and $P^+=\ker\eps\subseteq P$. Here $\tilde v$ denotes any representative of $v\in V$ in the larger space.  Applying this theorem in the above example gives
\[ V=\<b_\pm\>/\< b^2_\pm,b_0\>=\C\oplus\C,\quad \Delta_R b_\pm=b_\pm\tens t^{\pm 2}\]
which means that
\[ \Omega^1(\C_q[S^2])=\CE_{-2}\oplus \CE_{2}\]
as the direct sum of associated quantum bundles of monopole charges $\pm 2$. We identify them with the holomorphic and antiholomorphic differentials $\Omega^{1,0}$, $\Omega^{0,1}$ respectively in a double complex. This defines a $q$-deformed complex structure on the $q$-sphere, with several applications. Meanwhile, the invariant metric is
\[ g=q^2\extd b_-\bar\tens \extd b_++\extd b_+\bar\tens\extd b_--(2)_q\extd b_0\bar\tens \extd b_0\]
where $(2)_q=q+q^{-1}$, and the $q$-monopole connection induces a $\nabla$ which is torsion free and cotorsion free. It is given by
\[ \nabla \extd b_\pm=(2)_q b_\pm g,\quad \nabla \extd b_0=(1+(2)_q) b_0 g.\]
Its curvature is a multiple of ${\rm Vol}\tens \id$ on each of the two parts, where ${\rm Vol}=e_+\wedge e_-$ is the top form on the $q$-sphere. The Ricci curvature for the simplest choice of lifting map is
\[ {\rm Ricci}={q^{-1}(1+q^4)\over 2} g + (2)_q{(1-q^4)\over 2}i({\rm Vol}).\]
We see that Ricci acquires what in classical geometry would be an antisymmetric part as a result of the deformation. There is also a spinor bundle $\CS=\CE_{-1}\oplus\CE_1$ and a canonical Clifford map with resulting gravitational Dirac operator $\Dsl$. 

\medskip 
Since our machinery applies also to finite groups it would be nice to similarly have a finite quantum homogeneous space analogous to the above example, using finite groups and calculi on them. At the time of writing we lack a non-parallelizable such example. Finally, working with finite geometries obviously improves the divergences coming from the continuum, but so does (for example) $q$-deformation. For example, the natural categorically defined dimension of the left-regular representation 
\[ \dim_q(\C_q[SU_2])={\sum_{n\in\Z}q^{-{n^2\over 2}}\over 1-q^{-2}}\]
is finite for $q<1$. Similar formulae exist for all $\C_q[G]$ and have links to number theory and quantum mechanics in bounded domains \cite{MaSoi:ran}.

\section{Quantum gravity on finite sets}

So far we have shown that the general theory is quite able to construct canonical metrics and connections in a range of examples not possible in ordinary differential geometry. However, for quantum gravity we need to work with the entire moduli of geometries. This has been investigated in detail for small finite sets in \cite{MaRai:mod} and we discuss the results now. 

Here $A=\C(M)$ is the algebra of functions on a finite set $M$. A (necessarily noncommutative) differential structure is given by a subset $E\subset M\times M$ which we indicate as  `arrows' $x\to y$ for $x,y\in M$. Here $x\to x$ are not allowed. To keep things simple let us suppose that the calculus is bidirectional so if $x\to y$ then  $y\to x$. Even such unoriented graphs have not been classified, it is a wild problem. To simplify it let us stick to the parallelizable case where we assume that the graph is $n$-regular for a fixed $n$, i.e. for each $x$ the cardinality of the set of arrows out of $x$ is fixed. The simplest  class of vielbeins in this context boils down to a combinatorial part  and a continuous part at each point. The former is the choice at each $x$ of a numbering $1,\cdots, n$ of the arrows out of $x$, i.e., $x\to s_x(a)$ for some bijection $s_x$ at each $x$. The continuous degree of freedom is a non-vanishing normalisation which we denote $\lambda_a(x)$. The differential structure and vielbein are of the form
\[ \Omega^1(\C(M))=\C(M).\{e_a\ :\ a=1,\cdots n\},\quad e_a=\sum_{x\in M}\lambda_a(x)\delta_x\extd  \delta_{s_a(x)}\]
in terms of the differentials of Kronecker delta-functions.  The relations of the calculus are
\[ e_a f=f(s_\cdot(a))e_a,\quad \extd f=(\del^a f)e_a,\quad (\del^a f)(x)={f(s_x(a))-f(x)\over\lambda_a(x)}\]
where we see that the partial derivatives are finite differences. These calculi are all inner with $\Theta=\sum_a \lambda_a^{-1}e_a$. The finite group example in the previous section was of the above form with $s_x(a)=xa$ right translation in the group and $\lambda_a\equiv 1$. 
 
Again for simplicity we take the Euclidean frame metric so that the vielbein and covielbein are the same and $g=\sum_a e_a\bar\tens e_a$. We also have to fix a frame group $G$, its calculus and the action of its (braided) Lie algebra on the vielbein, all of which will generally be dictated by the geometrical picture we want to allow. If $n=2$ and we have in mind  a curve then a natural choice for $G$ is $\Z_2$ flipping the $e_1,e_2$ considered at angle $\pi$ (so the quantum cotangent bundle has an extra dimension). If for $n=2$ we have in mind a surface then a natural choice might be the group $\Z_{4}$ acting by $\pi/2$ frame rotations (if we consider $e_1,e_2$ at right angles), and so forth. We also have at some point to specify the $\wedge$ product for the higher exterior algebra through the choice of a $G$-invariant projector (this also gives the lifting $i$). In the simplest cases it could just be the usual antisymmetrization. In this way we fix the  `type' of data for the context of the model. 

To do quantum gravity with fixed cotangent dimension $n$ we then have to sum over all $m=|M|$ the number of points in the NCG, all $n$-regular graphs, all numberings $s_x$ on them, and integrate over all the functions $\lambda_a$ and all compatible connections $\alpha_i$ for the given framing and wedge product (probably we should sum over choices for the latter as well). If we take a (perhaps naive) path integral approach we should do all this with some action such as given by $R$, the Ricci scalar computed in the NCG after all the choices stated. So, a partition function
\[ \CZ=\sum_m \sum_{n-{\rm graphs}}\sum_{s}\int\prod_a\CD \lambda_a \int\prod_i\CD\alpha_i\ e^{-\beta \sum_{x\in M}R(x)}.\]

\begin{figure}
\[ \includegraphics{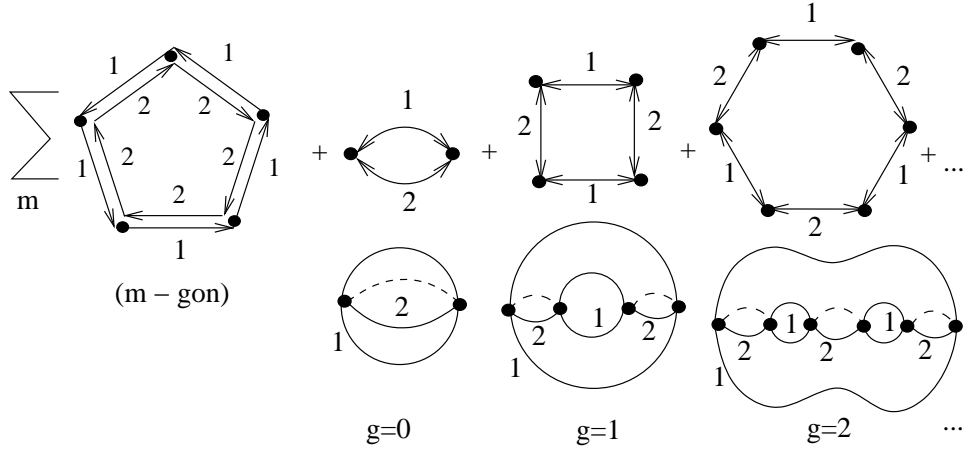}\]
\caption{Combinatorial part  of sum for  finite $n=2$ quantum gravity includes $2(g+1)$-gons as models of genus $g$ surfaces}
\end{figure}
Note that the choice $s$ is a {\em colouring} of the oriented graph into loops labelled from $\{1,\cdots,n\}$. This is because one can follow each number label from vertex to vertex; since the graph is finite it has to come back to itself at some point and every arrow must be coloured (in each direction). In addition to this the gravitational modes are allowed to set the physical length of each arrow which is the continuous part of the metric or vielbein data. For example, if we fix $n=2$ the graphs are polygons and the  colourings are either (i) colours 1,2 running oppositely around the polygon (ii) when $m$ is even; colours 1,2 alternating and with the two directions of each edge having the same colour. This gives the combinatorial part of the sum of configurations for $n=2$ quantum gravity  as shown schematically in Figure 1. The type (i) cases are the groups $\Z_m$ with their natural 2D calculus, which we view as approximations of a circle $S^1$. The type (ii) cases we propose correspond to surfaces of genus $(m-2)/2$ where each  $\leftrightarrow$ approximates a circle, so we have $m$ geodesic circles intersecting  at right angles  (since the labels alternate and describe orthogonal zweibeins), as indicated in Figure~1. (Different weights $\lambda_a$ will stretch these pictures about.) Thus the square of type (ii) is the natural geometry for $\Z_2\times\Z_2$ as a model of a torus, while the square of type (i) is  the natural geometry for $\Z_4$ as a model of $S^1$. We have a similar choice for $n=3$ and higher;  we can think of a trivalent graph as a triangulation of a surface (note the extra dimension, linked to the existence of  $\Theta$) or as the geometry of a 3-manifold. For example a cube with one kind of colouring can be a model of a sphere, or, with a different colouring, $\Z_2^3$ as a model of a 3-torus.  

Then for each combinatorial configuraton we have to integrate over the continuous degrees of freedom, preferably with a  source term in the action as we also want expectation values of operators, not just the partition function. Such an approach has been carried out in the $U(1)$ not gravitational case in \cite{Ma:cli}; the integrals over $\alpha$, while still divergent, are now ordinary not usual functional integrals and the theory is renormalisable. The gravitational case has not been carried so far but the moduli of $\alpha$ and its curvature have been investigated in the simplest $n=2$ cases\cite{MaRai:mod}. For example, consider the square with type (ii) colouring. We take zweibeins $e_1,e_2$ anticommuting as usual and $G=\Z_4$ with its 2D calculus generating $\pm \pi/2$ rotations in the group. The actions are therefore
\[ f^1\la e_1=e_2-e_1,\quad f^1\la e_2=-e_1-e_2,\quad f^2\la e_1=-e_2-e_1,\quad f^2\la e_2=e_1-e_2\]
(the Lie algebra elements are of the form $g-1$ in the group algebra, where $g$ is a group element).
To focus on the connections, let us fix $\lambda_a=1$, which is of course not the full story. Then $\extd e_a=0$ and the torsion-free equation becomes
\[ \alpha_1\wedge (e_2-e_1)=\alpha_2\wedge (e_2+e_1),\quad \alpha_1\wedge (e_1+e_2)= \alpha_2\wedge (e_1-e_2).\]
Let us also assume unitarity constraints in the form $e_a^*=e_a$ and $\alpha_1^*=\alpha_2$ (`antihermitian') which makes the cotorsion-free equation automatic as the conjugate of the torsion-free one. This is solved with $\alpha_i$ in basis $e_1,e_2$ given by functions $a,b$ subject to $\bar a=R_2a=-R_1a$, $\bar b=R_1b=-R_2b$, where $R_a$ are translations in each $\Z_2$ of $\Z_2\times\Z_2$. The resulting covariant derivative and Ricci scalar are
\[ \nabla e_1=(a e_1 + be_2)\bar\tens e_1+ (be_1-a e_2)\bar\tens e_2,\quad \nabla e_2=(- be_1+ae_2)\bar\tens e_1+(ae_1 + be_2)\bar\tens e_2\]
\[ R=-2(|a|^2+|b|^2),\quad \sum_x R(x)=-8 (|A|^2+|B|^2),\]
where $A,B$ are the values of $a,b$ are a basepoint (the values at the other 3 points being determined). The full picture when we include varying $\lambda_a$ has additional terms involving their derivatives and a regularity condition cross-coupling the $a,b$ systems (otherwise there is a kind of factorisation), see \cite{MaRai:mod} for details but with some different notations. Note that the model in the type (ii) case is very different from the simplicial interpretation in the type (i) models, and indeed we don't appear to have a Gauss-Bonnet theorem in the naive sense of $\sum_x R(x)=0$ for a torus. Also note that the torus case is too `commutative' for the weak Riemannian geometry to determine $\nabla$ from the metric and we see the larger moduli of connections. If we impose full $\nabla g=0$ then we do get $a,b$ determined from derivatives of $\lambda_a$ but this significantly constrains the allowed $\lambda_a$ and is not very natural. Our naive choice of $\Omega^2$ also constrains the $\lambda_a$, so the above is perhaps not the last word on this model.

 Looking at the diagrams in Figure~1 we see a similarity with matrix models and their sum over genus. Also note that the sum over graphs is not unlike Feynman rules for a $\phi^n$ interacting scaler theory in flat space. There vertices are the interactions, while graph arrows are propagators (doubled up in the type (i) colouring cf. in matrix models). As with Feynman rules, we follow the momentum (=colouring in our case) around in loops and sum over all values. The weights are different and given in our case by  NCG from the continuous degrees of freedom. In general terms, however, we see that finite set quantum gravity at fixed cotangent dimension $n$ has some kind of duality with flat space interacting $\phi^n$  scaler theory of some (strange) kind. There are likewise similarities with group field theory coming out of spin foams when expressed in a suitable way.
 
We also have obvious links with the causal poset approach to spacetime\cite{Sor}. The difference  is that NCG is a general framework that embraces both classical and these new approaches to geometry in a uniform manner, which addresses the main problem in say the poset as well as spin foam approaches. The main problem is to see how ordinary spacetime and geometry will emerge from such models. In NCG this is a matter of the algebra and the calculus. As $m\to \infty$ but with the calculus in a controlled limit, the finite differences will become usual differentials and, modulo some issues of extra dimensions,  we will see the classical geometry emerge within the uniform NCG framework. The specific comparison with causal sets is nevertheless interesting and suggests that a causal structure is `half' of a bidirectional differential structure in which  if $x\to y$ then the other way is not allowed. Given a poset we can double the arrows so that they are both ways, and proceed as above. Or indeed NCG also works with unidirectional calculi so we can work just  on the poset. One can also have a mixture of arrows, for example to model a black hole. In NCG one seeks a Hodge $*$-operator on the algebra of differential forms (it is known for all the examples above \cite{MaRai:ele,GomMa:non,Ma:sph}) after which one can do Maxwell and matter field theory as well as the gravitational case discussed above. 
 
\section{Outlook: Monoidal functors}

We have offered NCG as a more general framework definitely useful as an effective theory and conjecturally better as a foundation for geometry to avoid divergences. But is it the ulimate theory? In my view there is a deeper philosophical picture for quantum gravity that requires us to address the nature of physical reality itself, with NCG as a key stage of development. This was expounded many years ago in \cite{Ma:pri} so I shall be brief. Basically, the claim is that what is real are some measured outcomes $f(x)$ but whether $x$ is real and being measured by $f$ or $f$ is real and being measured by $x$ (what I have called observable-state duality) is like a gauge choice and not absolute as long as the choice is made coherently. This sets up a duality between different bits of physics, basically elevating Born reciprocity to a deeper conceptual level that applies not only to position and momentum but to geometry and quantum theory. In this setting Einstein's equations $G=8\pi T$ should be a self-duality equation identifying a quantum part of the theory with a classical geometry part, when both sides have been expressed in the same language, such NCG. Quantum group toy models \cite{Ma:pla}  demonstrate the idea in terms of Hopf algebra duality and  T-duality\cite{BegMa:poi},  where indeed a non-symmetric metric is inverted in the dual model (micro-macro duality).  Quantum group Fourier transform implements the Hopf algebra duality as explored for quantum spacetime models in our previous work, see \cite{Ma:algII} for a review.

But what about in the other direction, beyond Hopf algebras and NCG? In \cite{Ma:rep} we showed that the next self-dual type of object beyond Hopf algebras and admitting such duality was: a pair of monoidal categories $F:\CC\to \CV$ with a functor between then. To such a triple we constructed a dual $F^\circ:\CC^\circ\to \CV$ of the same type and a map $\CC\subset \CC^{\circ\circ}$. Therefore the next more complex theory to fulfill our self-duality requirement of  observer-observed symmetry after toy quantum group models should be a self-dual such object in some sense. There is a bit more going on here as unless $\CV$ is trivial (such as vector spaces) the duality operation extends the category (for example it takes a quantum group to its quantum double) so something should be projected out before we can speak about self-dual objects. That being said, it is interesting that 15 years later the notion of QFT on curved spaces has been nicely set-up  by Fredenhagen and coworkers precisely as a monoidal functor
\[F: \{{\rm Globally\ hyperbolic\ manifolds}\} \to C^*-{\rm Algebras}.\]
This is a noncommutative version of the functor $C(\ )$  that assigns to a manifold its commutative algebra of  functions, but is only half the story and not self-dual. However, deformations of $F$ could lead to a self-dual functor in the same way as one has genuine self-dual quantum groups. These would be systems with both quantum theory and gravity  along the lines of our experience with quantum group toy models. Moreover, the NCG constructions such as quantum bundles etc. extend to monoidal categories\cite{Ma:dia}, so this line can be explored.

\baselineskip 14pt
\bibliographystyle{unsrt}
\bibliography{biblio}

\end{document}